\documentclass[aps,twocolumn,showpacs]{revtex4-1}

\usepackage{graphicx}
\usepackage{amsmath}
\usepackage{amssymb}
\usepackage{braket}
\usepackage{color}
\usepackage{float,epstopdf}

\usepackage[colorlinks=true, citecolor=blue, urlcolor=blue ]{hyperref}

\input{epsf}
\begin{document}

%
%

\title{Beating no-go theorems by engineering defects in quantum spin models}
\author{Debasis Sadhukhan, Sudipto Singha Roy, Debraj Rakshit, Aditi Sen(De), and Ujjwal Sen}
\affiliation{Harish-Chandra Research Institute, Chhatnag Road, Jhunsi, Allahabad 211 019, India}

\date{\today}

\begin{abstract}

There exist diverse no-go theorems, ranging from no-cloning to monogamies of quantum correlations and Bell inequality violations, which restrict the processing of information  in the quantum world.
In a multipartite scenario, monogamy of Bell inequality violation and exclusion principle of dense coding are such theorems, which impede the ability of the system to have quantum advantage between all its parts. 
In ordered spin systems, 
the twin restrictions of translation invariance and monogamy of quantum correlations, in general, enforce the
bipartite states to be
neither Bell inequality violating nor dense-codeable.
We show that these quantum
characteristics, viz. Bell inequality violation and dense-codeability, can be resurrected, and thereby the no-go theorems overcome, by having quenched disorder in the system parameters leading to quantum spin glass or quantum random field models. 
We show that the quantum characteristics are regained even though the quenched averaging keeps the disordered spin chains translationally 
invariant at the physically relevant level of observables. 
The results show that it is possible to conquer constraints imposed by quantum mechanics in ordered systems by introducing impurities.




\end{abstract}


\maketitle

\section{Introduction}
\label{sec_introduction}


Disordered systems form one of the centerstages of studies in many-body physics~\cite{disorder-refs1,disorder-refs2,disorder-refs3}. Reasons include the fact that imperfection and impurities 
come naturally in several physical systems including ultracold atoms, nitrogen-vacancy centers in diamonds, solid-state systems, and ion traps. Moreover, such systems
exhibit exotic quantum phases and phenomena like Anderson localization \cite{disorder-refs1} and Bose glass \cite{Bose-glass}.
As a further example, during crystallization processes, atoms or molecules form solid structures  by arranging themselves in a way so that the total energy of the system gets minimized, but  thermal motion of the molecules, 
chemical impurities, or pressure differences, often lead to defects in their arrangements. As a consequence,  ideal mechanical, chemical, electronic, magnetic, or optical properties of the sample are expected to alter. 
Advances in experimental techniques have also made it possible to artificially 
realize disordered systems in different physical substrates~\cite{disorder-experiment}.

Defects, in general, reduce the physical properties like magnetization, conductivity, classical correlation, and quantum correlation \cite{HHHHRMP, arekta, Kavan-RMP} of the system~\cite{Jaworowski}. 
Thereby the system may loose its ability of performing in a better way than its classical counterpart. It has been reported that disorder reduces the fidelity of quantum state transmission as well as of
quantum gate implementation~\cite{Bayat}.

However, thermal fluctuation or impurities in the system, may lead to a counterintuitive enhancement of physical properties, known as ``disorder-induced order'' or ``order-from-disorder''.
In particular, observables like magnetization and classical correlations have been shown to increase in the presence of disorder in many-body systems~\cite{classical-corr}.
Quantum correlations, in general, are believed to be fragile quantities and can decay very fast in presence of impurities. Intuitively therefore, one expects that randomness, in general, will wipe out 
quantum correlations from the system. However, there are several examples for which this intution fails \cite{qc-more-in-disorder}.

Quantum mechanics places strict restrictions in the form of ``no-go theorems'', like no-cloning~\cite{wootersnc} and monogamy of quantum correlations~\cite{monogamy}, on information processing tasks (see also
\cite{no-go-etc}). In this paper, 
we concentrate on two restrictions imposed by the quantum mechanical principles -- monogamy of Bell inequality violation~\cite{bell-monogamy} and exclusion principle of  classical information transmission~\cite{prabhu}. 
In a multipartite scenario, with a boss and several subordinates, the laws  slate that if the shared quantum state between the boss and a single subordinate exhibits quantumness, either by violating Bell inequality or by
being dense codeable, then the other channels between the boss and the subordinates are prohibited from possessing  the same quantum advantage, and hence, enforces limitations upon the quantum information processing tasks 
possible in that scenario. 

It is easy to see therefrom that the two-qubit states obtained from translationally invariant systems, which include the ground states of one-dimensional translation-invariant quantum spin models (without disorder), 
neither violate Bell inequality and nor have quantum advantage in dense coding \cite{bhalo-paper}. 
The two-pronged restriction imposed by monogamy and translation invariance causes \emph{all} two-qubit states 
of such multiparty systems
to be devoid of the quantum advantages. 
The same arguments are true for an arbitrary isotropic higher-dimensional lattice.
We now ask the following question: Is it possible to regain the quantum advantages in these two-qubit states in some physical many-body system, 
while still retaining the translation invariance of the system, at least at the level of observables
under study, i.e. at the level of the amount of Bell-inequality violation and the capacity of dense coding? We answer the question in the affirmative by using quenched disordered spin systems.


We consider quenched disordered one-dimensional quantum spin-1/2 systems. We show that even though translation symmetry is present in these systems after quenched averaging, such disordered models can overcome the 
hurdle of Bell monogamy and exclusion principle of dense coding. 
First, we  show that in the disordered quantum $XY$ spin glass and in the random field quantum $XY$ model, the quenched averaged quantities
for the amount of Bell inequality violation as well as the capacity of dense coding, of the nearest-neighbor zero-temperature state can attain nonclassical values and thereby overcome the monogamy 
constraints,
despite the fact that the post-quenched 
averaged quantities are translation-invariant. The analysis is carried out by applying the Jordan-Wigner transformation to the disordered $XY$ models~\cite{LSM, barouch1, barouch2}. The phenomena observed them is potentially generic, 
in that we have also demonstrated in models which do not lend themselves to an analytical treatments.
Precisely, we extend the analysis and demonstrate the phenomena for quenched disordered  quantum 
Heisenberg spin glasses, for which the investigation is performed via the density matrix renormalization group (DMRG) technique. 
Finally, we carry out the finite-size scaling analyses for both the quenched observables in all the models considered. 


The paper is organized as follows. In Sec.~\ref{sec:monogamy}, we review the monogamy of Bell inequality violation and the exclusion principle of dense coding.  
The results are presented in Secs.~\ref{sec: XY} and~\ref{sec:XYZ}. Sec.~\ref{sec: XY} introduces the disordered $XY$ spin models and briefly discusses the methodology, following which it presents the results which 
show how quantum systems, possessing imperfections, lead to clear advantages over the corresponding clean systems. The results for the quenched disordered Heisenberg spin glass are presented in Sec.~\ref{sec:XYZ}. 
Finally, we conclude in Sec.~\ref{sec:conclusion}.

\section{Monogamy of Bell inequality and dense coding capacity}
\label{sec:monogamy}

   It is known from the celebrated Bell theorem \cite{Bell} that the violation of the Bell inequality by a two-party state
   guarantees that the state cannot have a 
local realist description.
Given any two-qubit state, $\rho$, violation of the Bell-CHSH~\cite{chsh} inequality demands 
\begin{equation}
  \langle{\cal B}_{CHSH}\rangle_{\rho} >2  \label{eqn:bell}.
\end{equation} 
Let us define $U=T_{\rho}^T T_{\rho}$, where $T_{\rho}^{mn}=Tr(\sigma_m \otimes\sigma_n\rho)$ are the elements of the corresponding correlation matrix, $T_{\rho}$. It can be shown that after maximization over the 
local measurements, we have 
\cite{bell-Horodecki}
\begin{equation} 
\max  \langle{\cal B}_{CHSH}\rangle_{\rho} \equiv \langle{\cal B}_{max}\rangle_{\rho}= 2\sqrt{M(\rho)},
\end{equation}
where 
$M(\rho)=u_1+u_2$, with $u_1$ and $u_2$ being the two largest or the largest and the second-largest eigenvalues of $U$. In order to violate the Bell-CHSH inequality, one therefore requires
\begin{equation}\label{eqn:Mrho}
 M(\rho)>1. 
\end{equation}
In case of multipartite states, \textit{if the quantum state shared by any two subparts of a multiparty system leads to a Bell inequality violation, then it precludes 
its violation for the states which the subparts share with the other parts of the total system.} This is referred to as  monogamy for Bell inequality violation for the multiparty quantum states~\cite{bell-monogamy}.

We define a quantity $\delta(\rho_{AB})=\max\{0,M(\rho_{AB})-1\}$, which  quantifies the amount of Bell inequality violation for the two-qubit states, and  in the following sections,
we will investigate its behavior while exploring different physical many-body systems.

On the other hand, the quantum dense coding protocol incorporates a sender-receiver scheme for communicating classical information over a quantum channel. If we consider that our conventional sender, Alice, and receiver, Bob, initially 
share a state $\rho_{AB}$, with $d_A$ and $d_B$ being the dimensions of the Hilbert space corresponding to Alice's and Bob's parts respectively, then the  dense coding capacity turns out to be~\cite{plenio}
\begin{equation}\label{eqn:dense}
 \mathcal{C}(\rho_{AB})= \log_2 d_A + C^{adv}(\rho_{AB})
\end{equation}
bits.
The quantity $C^{adv}(\rho_{AB})=\max\{0,  S(\rho_A)-S(\rho_{AB})\}$ is referred to as the ``quantum advantage'' of dense coding of the state $\rho_{AB}$ over the classical channel \cite{vonneumann}. 
This is justified by the fact that $\log_2{d_A}$ bits of classical information can be transmitted by sending a $d_A$-dimensional quantum system. A bipartite quantum state is said to be dense codeable if it has a 
positive quantum advantage of dense coding. In a multipartite scenario, the ``exclusion principle" for quantum dense coding demands that 
\textit{if any two subsystem of a multiparty quantum system shares a dense codeable state, then they can't share any 
such quantum state efficient for dense coding, simultaneously, with other parts of the system}.

Let us illustrate the above no-go theorems for a three-party state. 
When a tripartite state \(\rho_{ABC}\) is shared between A, B, and C, monogamy of Bell inequality violation and exclusion principle implies that if the reduced state \(\rho_{AB}\) violates local realism or has quantum advantage in dense coding, 
i.e. if 
\(\delta(\rho_{AB})\) or \(C^{adv}(\rho_{AB})\) is positive, then the reduced state at AC will have \(\delta(\rho_{AC}) = 0\) or \(C^{adv}(\rho_{AC})=0\)  respectively.

\section{Advantages in Disordered Quantum $XY$ models}
\label{sec: XY}

In this section, we consider the disordered quantum $XY$ models for testing the monogamy of Bell violation and the exclusion principle for dense coding capacity. 
We begin by briefly discussing the idea of quenched averaging in the following subsection.
 
\subsection{Quenched averaging}
In the present work, for all purposes, the type of disorder that has been used is ``quenched".
Spin glass states are those which emerge due to the presence of such type of disorder in the system and the term ``glass'' comes from the analogy with the chemical glass which is formed by quenching a liquid. 
The term ``quenched" signifies that the time over which the dynamics of the system takes place is much smaller than the time scale over which there is a change in a particular realization of parameters governing the disorder in the system.
This leads to the fact that while calculating the quenched averaged value of a physical quantity, we need to perform the averaging of several expectation values of that quantity, each of which is obtained for a fixed configuration, over the 
relevant probability distribution of the configurations. 

\subsection{The model and the methodology}
The Hamiltonian for the one-dimensional disordered quantum $XY$ spin chain in a random transverse field is given by
\begin{align} 
H =   \kappa \Big[ \sum_{i = 1}^{N} \frac {J_i}{4} \Big((1+\gamma) \sigma_i^x \sigma_{i+1}^{x} + (1 - \gamma ) \sigma_i^y \sigma_{i+1}^{y}\Big) \nonumber \\
-  \sum_{i=1}^N \frac {h_i}{2} \sigma_i^z \Big],
\label{eqn: XY}
\end{align}
where $\kappa J_i$ is the coupling strength between the $i^\text{th}$ and $(i+1)^\text{th}$ site, $\kappa h_i$ represents the field strength at the $i^\text{th}$ site,  
and $\gamma$ is the anisotropy constant. $\kappa$ is a constant and has the unit of energy, while $J_i,h_i,$ and $\gamma$ are dimensionless. Here, $\sigma^j,$ for $j=x,y,z,$ correspond to the Pauli spin matrices. 
For the ordered system, all the $J_i$ and $h_i$ are separately equal, and are denoted by $J$ and $h$, respectively.  Here we have assumed cyclic boundary condition, so that the $(N+1)^{th}$ and the $1^{st}$ sites are equivalent. 

The ordered model is exactly solvable via successive use of the Jordan-Wigner, Fourier, and  Bogoliubov transformations~\cite{LSM,  barouch1, barouch2}, while the disordered model is not. However,  
the same procedure can again lead us to 
the one- and two-site reduced density matrices for the disorder case.
For completeness, we briefly review the mechanism here.
First, we map the Pauli spin operators to the spinless fermions via the Jordan-Wigner transformation, so that Eq.~(\ref{eqn: XY}) becomes
\begin{align}
H = \kappa\Big[\sum_{i,j = 1}^{N} c_i^{\dagger}A_{ij}c_j + \frac 12\sum_{i,j=1}^{N} \big( c_i^{\dagger}B_{ij}c_{j+1}^{\dagger} + h.c. \big)\Big],
\label{eqn: XYAB}
\end{align}
where $A$ and $B$ are symmetric and antisymmetric real $N \times N$ matrices, respectively, and are given by 
\begin{align}
A_{ij} &= -h_i\delta_{ij} + \frac{J_i}{2} \delta_{i+1,j} + \frac{J_{j}}{2} \delta_{i,j+1}, \nonumber\\
B_{ij} &= \frac{\gamma}{2} (J_i \delta_{i+1,j} - J_j \delta_{i, j+1}), \nonumber
\end{align}
with $A_{1N}=A_{N1}=J_N$ and $ B_{1N} = - B_{N1} = - \frac{\gamma}{2}J_N $ for the cyclic boundary condition. Here, the $c_i^{\dagger},c_i$ are spinless fermionic operators obtained via the Jordan-Wigner transformation. Defining $\Phi^T_{k}$ via the eigen-equation
\begin{align}
(A-B)(A+B)\Phi_k^T = \Lambda_k^2\Phi_k^T, \label{eqn: EG}
\end{align}
with eigenvalue $\Lambda_k$ and obtaining the corresponding $\Psi_k$  from the equation
\begin{align}
\Psi_k^T = \Lambda_k^{-1} (A+B) \Phi_k^T,
\end{align}
we can calculate the correlation matrix $G$, defined as
\begin{align}
G_{ij} = -\sum_{k} \psi_{ki} \phi_{kj} = -({\bf \Psi}^T{\bf \Phi})_{ij},
\end{align}
where ${\bf \Phi}$ and ${\bf \Psi}$ are the matrices $\phi_{ki}$ and $\psi_{ki}$, with $\phi_{ki}$ ($\psi_{ki}$) being the \(i\)th element of $\Phi_{k}$ ($\Psi_{k}$). Finally, one can show that the magnetizations and and two-point correlation functions of the zero-temperature state can be easily obtained from the correlation matrix $G$. We get $m_i^z = -G_{ii}$ and $m_i^x = m_i^y = 0$.  The diagonal correlations are given by
\begin{align}
T^{xx}_{i,i+1} &= G_{i,i+1} \label{eqn: Txxf},\\
T^{xx}_{i,i+1} &= -G_{i+1,i} \label{eqn: Tyyf},\\
T^{zz}_{i,i+1} &= G_{i,i}G_{i+1,i+1} - G_{i,i+1}G_{i+1,i},
\label{eqn: Tzzf}
\end{align}
while all off-diagonal correlations vanish. The one- and the two-site density matrices can now be easily constructed from the one- and two-point correlation functions and consequently, the Bell inequality violation (Eq.~(\ref{eqn:Mrho})) and the dense coding capacity (Eq.~(\ref{eqn:dense})) can be computed.

\subsection{Anisotropic $XY$ Spin Glass}
 
Let us now consider the quantum $XY$ model of $N$ spins interacting via site-dependent nearest neighbour exchange interactions, $J_i$, which are identically and independently distributed $(i.i.d.)$ with
Gaussian probability distribution, while the field strength, \(h_i\), at each lattice site is kept constant. The corresponding Hamiltonian follows from Eq.~(\ref{eqn: XY}) by setting $h_i=h$ for all $i=1,2,\cdots,N$ and by letting $J_i$ to follow 
the probability distribution  
\begin{equation}
\label{eqn: distribution}
 P(J_i)=\frac{1}{\sqrt{2\pi\sigma}} \exp\Big[-\frac{1}{2}\Big(\frac{J_i-\bar{J}}{\sigma}\Big)^2\Big],
\end{equation}
where $\bar{J}$ and $\sigma$ are, respectively, the mean and the standard deviation of the distribution. 



For the ordered case, due to translational symmetry, all the nearest-neighbor density matrices are equal, and hence the monogamy of Bell correlations restrains any two-party reduced density matrix of the $N$-party zero-temperature state 
of the $XY$ Hamiltonian from violating the Bell-CHSH inequality. 
Similarly, the exclusion principle for dense coding implies that the maximum achievable dense coding capacity for a quantum state shared between any two-party state of the 
\(N\)-party system, is restricted to the classical dense coding capacity, 
$\log_2{d}$, where $d$ is the dimension of the Hilbert space corresponding to the sender, and hence no quantum advantage is possible in any two-party reduced state. 
These are consequences of the monogamy relations mentioned in Sec.~\ref{sec:monogamy}, as applied to the 
case of the ordered chain. 

More precisely, let $\rho_{AB}$ be a two-party state reduced from the zero-temperature state in the ordered case. Here A and B are disjoint collections of lattice sites of the one-dimensional chain. The 
translation invariance of the ordered chain (with periodic boundary condition for finite $N$) implies that we can always find a collection C of 
lattices that is disjoint from both A and B such that the reduced states $\rho_{AB}$ and $\rho_{BC}$ of the zero-temperature state are equal. The subtle assumption here is that the chain is  sufficiently large, so that C 
does not overlap with A or B (cf. \cite{bhalo-paper}). In particular if $\rho_{AB}$ is a nearest-neighbor (two-site) density matrix, we only need $N \geq 3$. Applying now the monogamy relations to the state $\rho_{ABC}$, we obtain
the  stated results 
for the ordered case in the translational invariant scenario. While we have considered only the one-dimensional case explicitly in this paper, the same arguments hold for any isotropic higher-dimensional lattices.

For the spin glass system, the above line of argument cannot be applied. The zero-temperature properties of the system are physically 
relevant only after quenched averaging has been performed, and post-quenching, these properties are again translationally invariant, just like for the ordered case. 
And so ${\cal Q}^{AB}_{\lambda}={\cal Q}^{BC}_{\lambda}$ for any physical property, ${\cal Q}$, for the reduced states at AB and BC. 
Note that throughout this paper, we associate the subscript $\lambda$ to a quantity, if it is quenched averaged.
The ${\cal Q}^{AB}_{\lambda}$ and ${\cal Q}^{AC}_{\lambda}$, however, do not 
correspond to a single state of ABC, and so the monogamy argument of the ordered case do not carry over to the disordered ones. We are therefore confronted with the possibility that disordered systems can give rise 
to situations, which
despite being translationally invariant, 
will have nearest-neighbor Bell inequality violation and quantum advantage in dense coding. Whether this is actually the case, however requires explicit investigations.

\begin{figure}[h!]
\includegraphics[angle=0,width=8.25cm]{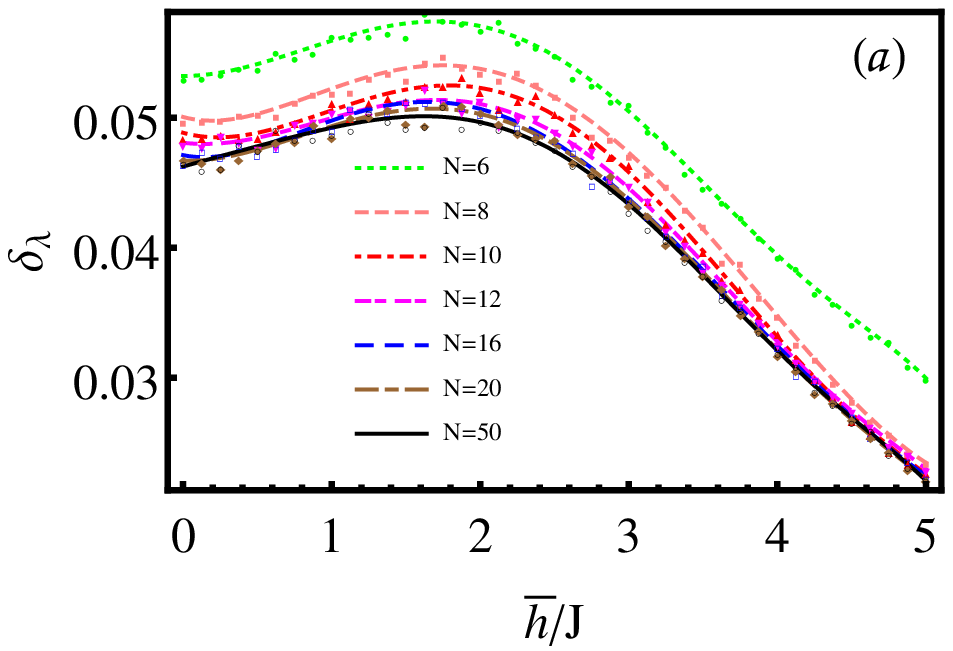}\label{fig: BVXYJ} \\
\includegraphics[angle=0,width=8.25cm]{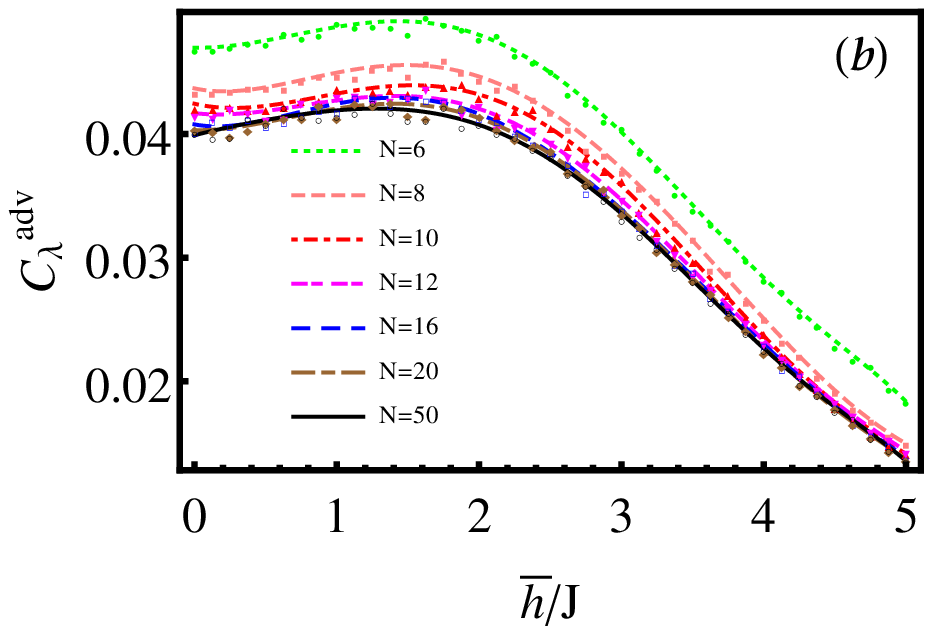} \label{fig: DCXYJ}
\caption{(Color online.) Overcoming monogamy in quantum XY spin glass. (a) Plot of quenched averaged Bell inequality violation ($\delta_{\lambda}$) on the vertical axis against $\bar{J}/h$ on the horizontal axis for nearest-neighbor
spins of the zero temperature state in the quantum anisotropic $XY$ spin glass for different $N$. Here we have chosen the uniform field strength $h=0.4$, the anisotropy constant $\gamma=0.5$, and the disorder strength $\sigma=1.0$. (b) 
This is the same as (a), except that the quenched averaged quantum advantage for dense coding, $C^{adv}_{\lambda}$ is plotted on the vertical axis. All quantities are dimensionless, except $C^{adv}_{\lambda}$, which is in bits.}
\label{fig: XYJ}
\end{figure}

In Fig.~\ref{fig: XYJ}, we show the quenched averaged Bell inequality violation, $\delta_{\lambda}$ (Fig.~\ref{fig: XYJ}(a)), and the quenched averaged quantum advantage of dense coding, $C^{adv}_{\lambda}$ (Fig.~\ref{fig: XYJ}(b)), 
as a function of $\bar{J}/{h}$, for any pair of nearest-neighbor spins of the zero-temperature state.  Fig.~\ref{fig: XYJ} clearly shows that 
after quenched averaging, the system violates the Bell-CHSH inequality and has quantum advantage in dense coding for the nearest-neighbor spins, which is a part of 
any nearest-neighbor three-party state of the \(N\)-party state. Enhancement of a system characteristic by the introduction of disorder is a well-known phenomenon 
and has a long history in the literature. This has been referred to as ``order from disorder", ``disordered-induced order", ``disorder induced enhancement", etc. \cite{classical-corr, qc-more-in-disorder}. Here we find that the
introduction of disorder can even lead to a qualitative change in the system character -- a hitherto (i.e., in the ordered case) monogamous situation can turn into a monogamy violating one. The violation is in an extreme 
sense, since both the two-party reduced states of the three-party cluster violate Bell inequality with equal strength. The same is true for dense-codeability.

Below, we will see that the phenomena are far from being specific to the system considered. They appear rather generically in  several paradigmatic physical systems. 

 
 \subsection{Random Field Quantum $XY$ Model}
 
\begin{figure}[t]
\includegraphics[angle=0,width=8.25cm]{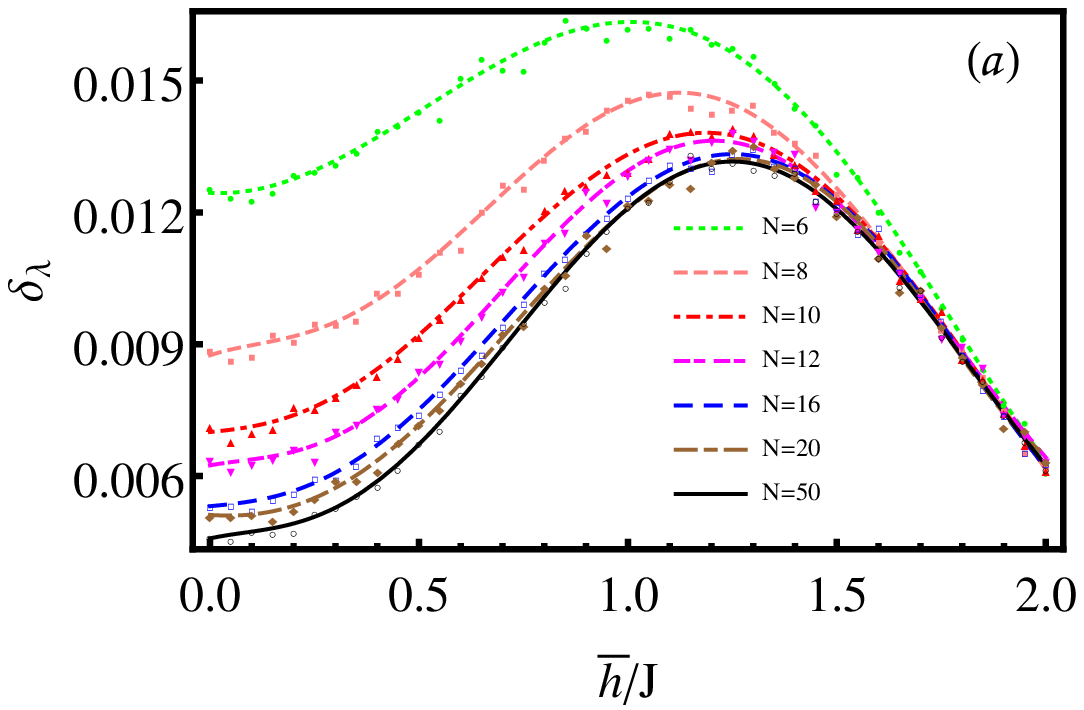} \label{fig: BVXYh}\\
\includegraphics[angle=0,width=8.25cm]{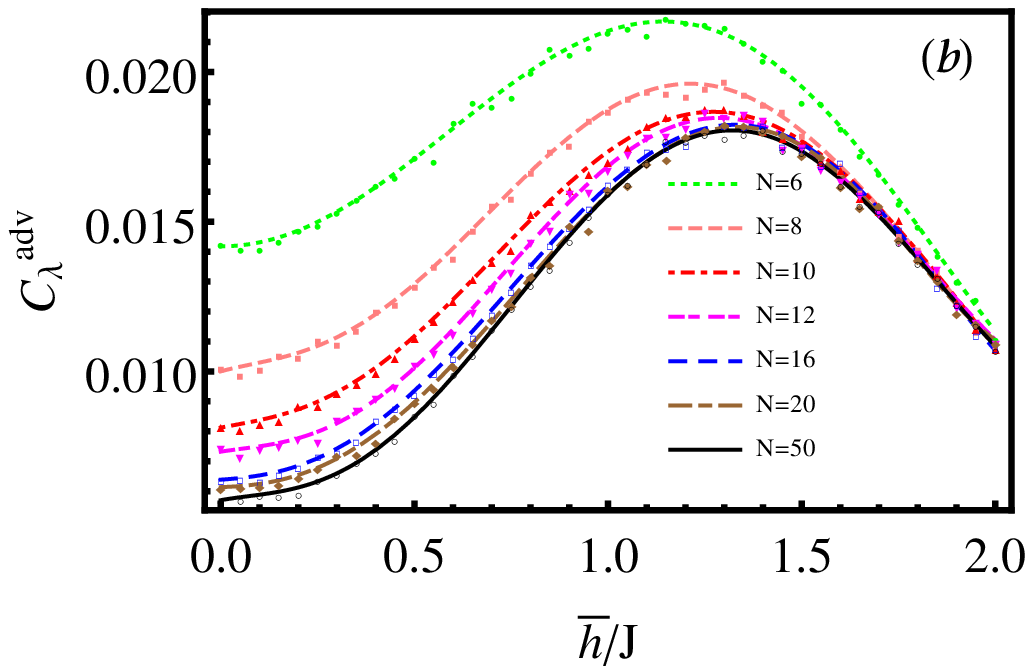} \label{fig: DCXYh}
\caption{ (Color online.) Overcoming monogamy in random field quantum $XY$ model. The considerations here are exactly the same as in Fig.~1, except that we are here using the random field quantum $XY$ model with the $h_i$ being $i.i.d$ Gaussian random variables (mean $\bar{h}$ and unit standard deviation), and then $J=1$.}
\label{fig: XYh}
\end{figure}

We now introduce the randomness in the field while keeping the coupling strength uniform. The Hamiltonian follows from Eq.~(\ref{eqn: XY}) by setting $J_i=J$ for $i=1,2,\cdots,N$ and by requiring the quenched random 
parameter $h_i$ to follow the Gaussian distribution with mean $\bar{h}$ and standard deviation $\sigma$. We obtain thereby the random field quantum $XY$ model, whose Hamiltonian is given by 
 \begin{align}
H^{RF}=\kappa\Big[ \frac{J}{4}\sum_{i}^N  [(1 + \gamma)\sigma_i^x\sigma_{i + 1}^x+(1 - \gamma)\sigma_i^y\sigma_{i + 1}^y] \nonumber \\
-\sum_{i}^N\frac{h_i}{2}\sigma_i^z \label{eqn: XYDh}\Big],
\end{align}
where $\kappa$ has the unit of energy, while $J$, $\gamma$, and $h_i$ are dimensionless.
Using the techniques similar to the ones in the preceding subsection, we find out the single- and two-site correlations and hence the Bell inequality violation and the advantage in dense coding. 
Fig.~\ref{fig: XYh} shows the variation of the quantities $\delta_{\lambda}$ (Fig.~\ref{fig: XYh}(a))  and $C^{adv}_{\lambda}$ (Fig.~\ref{fig: XYh}(b)) with respect to $\bar{h}/J$. Similar to the 
case of the disordered $XY$ spin glass, we find quantum advantage in both the quantities for the set of parameters considered here. Thus, here too, introduction of the quenched disorder in the system helps to 
overcome the restriction put by the monogamy relations.

\subsection{Scaling of $\delta_{\lambda}$ and $C^{adv}_{\lambda}$ for $XY$ Model}

\begin{figure}[h!!]
\includegraphics[angle=0,width=4.2cm]{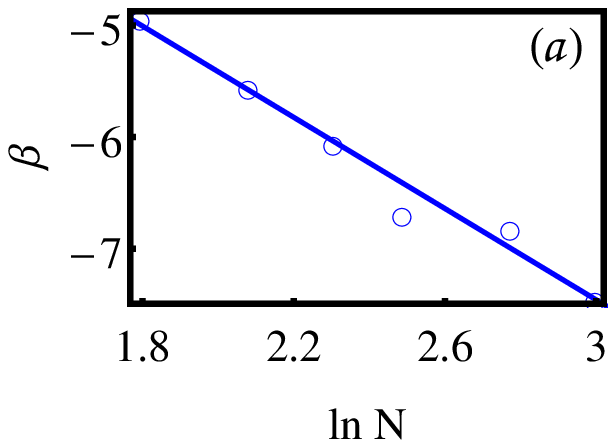} \label{fig: BVScalingJ}
\includegraphics[angle=0,width=4.2cm]{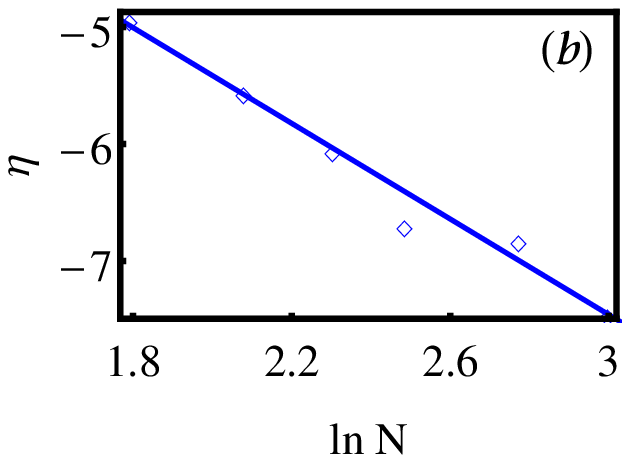} \label{fig: DCScalingJ}\\
\includegraphics[angle=0,width=4.2cm]{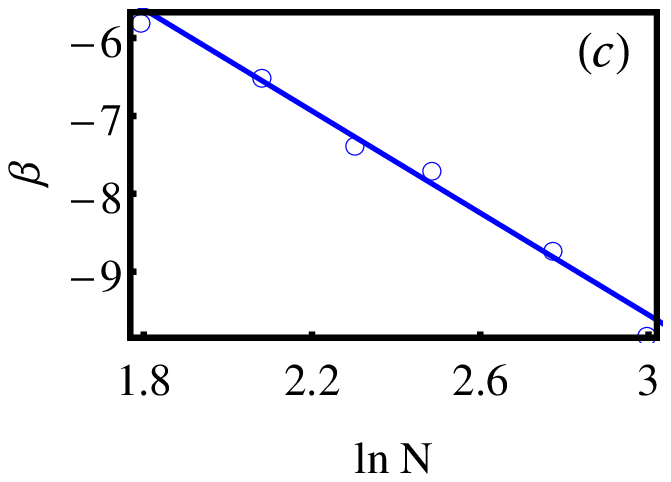} \label{fig: BVScalingh}
\includegraphics[angle=0,width=4.2cm]{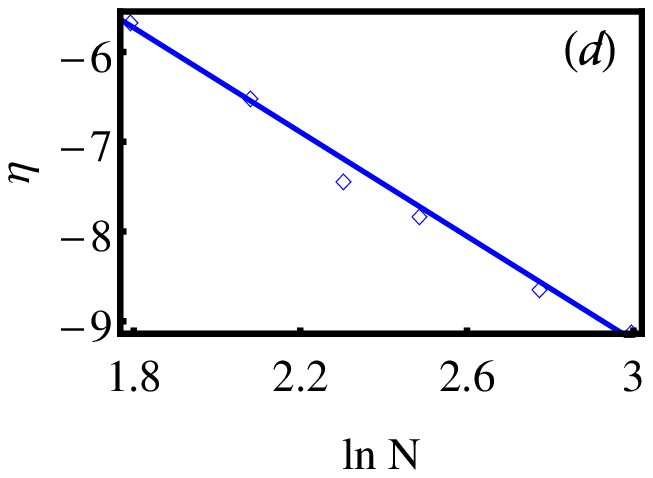} \label{fig: DCScalingh}
\caption{(Color online.)
Finite-size scaling analysis for Bell inequality violation and advantage in dense coding for the disordered quantum $XY$ models. The vertical axes represent $\beta$ (in panels $(a)$ and $(c)$, using circles) and 
$\eta$ (in panels $(b)$ and $(d)$, using diamonds), where $\beta=\log \big|\delta_{\lambda, max}(N)-\delta_{\lambda, max}(N_c) \big|$ and  $\eta=\log \big| C^{adv}_{\lambda, max}(N)-C^{adv}_{\lambda, max}(N_c) \big|$,
for a chain of length \(N\). We treat $N=N_c \equiv 50$ as the infinite chain.  The solid lines are the best linear fits. 
While the panels (a) and (c) are for the quantum \(XY\) spin glass, 
the panels (b) and (d) are for the random field \(XY\) model.
The quantity \(b\) is dimensionless, \(c\) is measured in \(\ln (\mbox{bits})\), and the horizontal axes are measured in \(\ln\) of the number of spins.}
\label{fig: Scalingh}
\end{figure}

From Figs.~\ref{fig: XYJ} and  \ref{fig: XYh}, it is clear that variation of both the quantities, $\delta_{\lambda}$ and $C^{adv}_{\lambda}$, for large systems, 
mimic the pattern obtained for N=20. Hence, the systems with $N>20$ spins can safely be assumed to serve the purpose of infinite spin chains. For our calculations, we use the value for \(N=50\) as for the infinite system.
For the \(XY\) spin glass,
we plot $\ln \big|\delta_{\lambda,max}(N)-\delta_{\lambda, max}(N_c) \big|$ and $\ln \big| C^{adv}_{\lambda, max}(N)-C^{adv}_{\lambda, max}(N_c) \big|$ against \(\ln N\) in 
Figs.~\ref{fig: Scalingh}(a) and \ref{fig: Scalingh}(b), respectively.  We find that $\delta_{\lambda, max}$ and $C^{adv}_{\lambda, max}$ decay as $N^{-2.05}$ and $N^{-2.30}$, respectively. 
Here the subscript ``$max$''
indicates 
that the scaling is done for the maximum values of both the quantities. Figs.~\ref{fig: Scalingh}(c) and \ref{fig: Scalingh}(d) show the variation of the same quantities for the random field \(XY\) model. 
We find that $\delta_{\lambda,max}$ and
$C^{adv}_{\lambda, 
max}$, for this case, scale as  $N^{-3.27}$ and $N^{-2.90}$, respectively.
The scaling analysis and the overall behavior of the quantities with increasing \(N\) clearly indicate that the violation of monogamy and the quantum advantage of classical information transmission 
will be sustained even in the thermodynamic limit, since for \(N>20\), the overall behavior of the physical quantities do not change with the increase of \(N\), within the accuracy considered.

\section{ Advantage in quantum $XYZ$ spin glass}
\label{sec:XYZ}
In previous section, we considered quenched disordered quantum $XY$ models. It was shown that such disorder driven systems are endowed with quantum advantage as compared to the clean systems with no quantum advantage, 
for certain physical properties. The ordered quantum $XY$ model is exactly solvable. The corresponding quenched disordered systems are also analytically tractable up to a certain extent. To find the extent to which the phenomenon 
considered here is generic, we also consider a non-integrable model, viz. the quenched disordered quantum $XYZ$ spin glass. We find that the monogamy for Bell inequality and exclusion principle for dense coding are again violated in this 
Heisenberg system, in a stronger way than in the random \(XY\) models. Indeed, the violations here (for suitable choice of system parameters) are an order of magnitude higher than in the disordered \(XY\) cases.

The one-dimensional quantum $XYZ$ Heisenberg Hamiltonian with random nearest-neighbor couplings in a magnetic field is given by
\begin{eqnarray} \label{eqn: XYZ}
 H = \kappa\Big[\sum_{i=1}^{N-1}\Big[J_{i}\big[(1+\gamma)\sigma_i^x\sigma_{i+1}^x+(1-\gamma)\sigma_i^y\sigma_{i+1}^y\big]&+ \nonumber \\
\Delta \sigma_i^z\sigma_{i+1}^z\Big] 
-h\sum_i \sigma_i^z \Big],&
\end{eqnarray}
where $\kappa \Delta$ is the coupling strength along the $z$-direction.
\begin{figure}[t]
\includegraphics[angle=0,width=8.6cm]{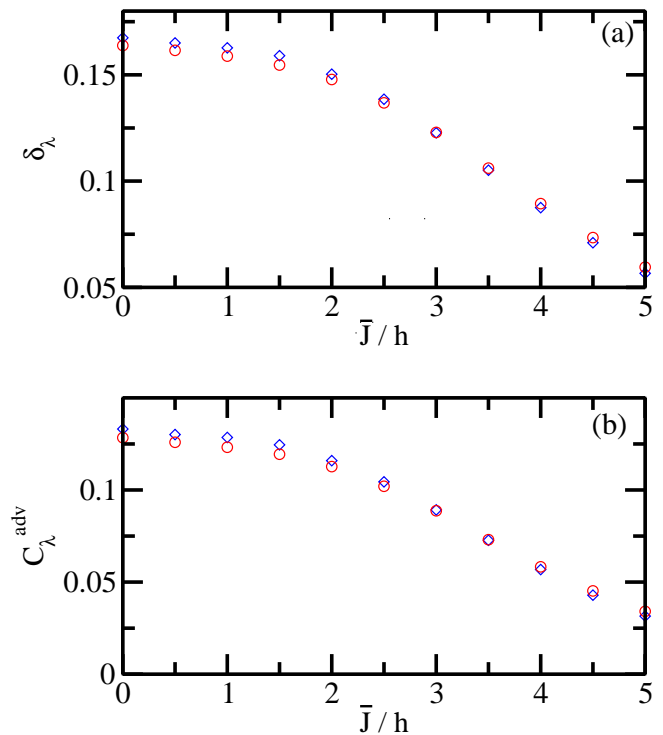}
\caption{ (Color online.) Bell inequality violation and quantum advantage in dense coding for the $XYZ$ spin glass. Illustration of $\delta_{\lambda}$ (in panel (a))
and $C^{adv}_{\lambda}$ (in panel (b)) between the spin-pairs $(N/2-1, N/2)$ (blue diamonds) and $(N/2, N/2+1)$ (red circls)  for varying $\bar{J}/h$. The data is obtained for $N=20$, using DMRG. For all the cases, the parameters $\gamma$, 
the $\Delta$ and the $\sigma$ are kept constant at $0.5, 0.7$, and $1.0$, respectively. 
All quantities are dimensionless, except $C^{adv}_{\lambda}$, which is in bits.} 
\label{fig:periodic}
\end{figure}
Here the $i.i.d$ random variables $J_i$ follow the distribution pattern given in Eq.~(\ref{eqn: distribution}). We have restricted our area of interest to the ground state physics only.

\begin{figure}[t]
\includegraphics[angle=0,width=7.8cm]{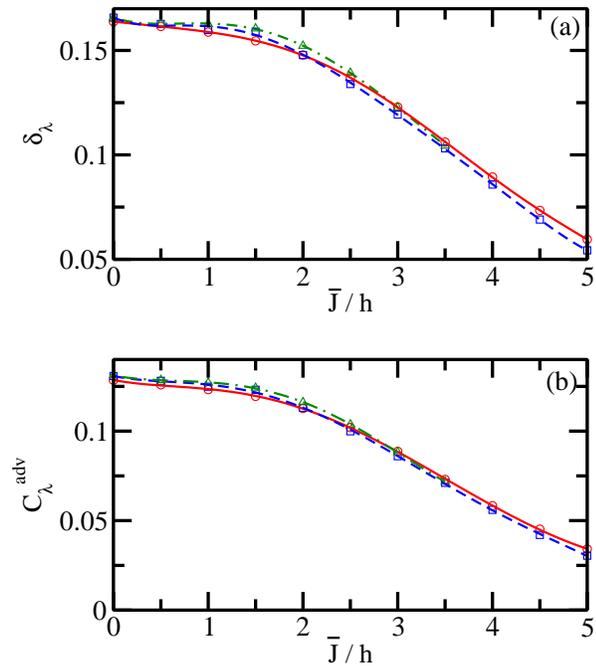}
\caption{(Color online.) $\delta_{\lambda}$ (in panel (a)) and $C^{adv}_{\lambda}$ (in panel (b)) as functions of $\bar{J}/h$ between the spins $N/2$ and $N/2+1$. Circles, squares, and triangles represent
the results for $N=20, 30$ and $50$, respectively. In both panels (a) and (b), solid lines show polynomial fits to the data points. 
The parameters and notations are the same as in Fig. \ref{fig:periodic}.
}
\label{fig:scaling}
\end{figure}

\begin{figure}[h!!]
\includegraphics[angle=0,width=8.0cm] {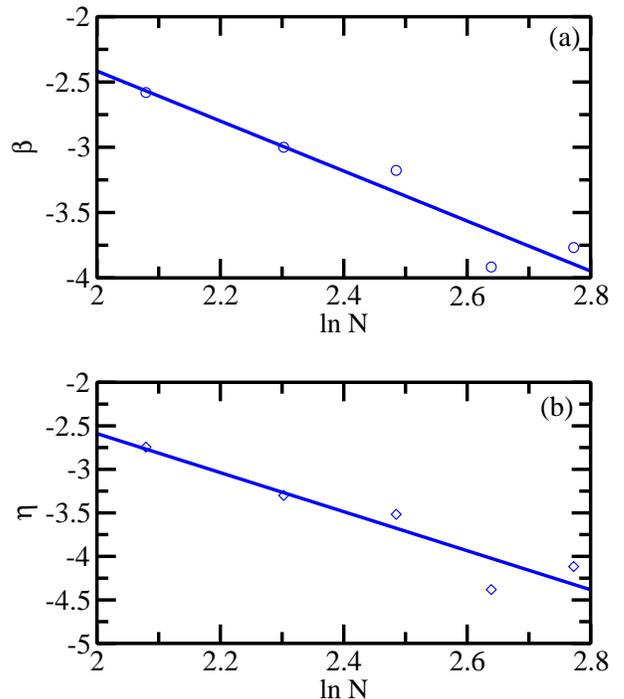}
\caption{ (Color online.) Finite-size scaling analysis for Bell inequality violation and advantage in dense coding for the quantum \(XYZ\) spin glass. 
The circles and the diamonds represent $\beta$ (in panel(a)) and $\eta$ (in panel(b)) respectively.
All notations are the same as in Fig. \ref{fig: Scalingh}.
} 
\label{fig:scaling1}
\end{figure}

In order to investigate the monogamy relations for Bell inequality violation as well as exclusion principle for dense coding, the ground state for the system characterized  by the Hamiltonian in Eq.~(\ref{eqn: XYZ}) is 
obtained by the numerical technique called the DMRG method \cite{DMRGpaper1}. First, the infinite-size DMRG method is performed  iteratively, where at each iteration the dimension of the system is reduced by 
selectively choosing the
most relevant basis states important for describing the system while truncating the rest. Afterwards, several finite size DMRG are also carried out on the disordered chain in order to increase the accuracy. The quenched averaged 
values of the physical quantities are obtained by averaging over 5000-8000 random realizations.   

DMRG gives much less accurate results in the case of periodic boundary conditions. However, the advantage of the open boundary condition comes at the expense of the boundary effects. Nevertheless, 
an adequate description of the Bell monogamy and the exclusion principle for dense coding is possible provided the system size is not too small and the measurement of the observables on either fringe are excluded.
In order to forego the boundary effect, we focus on the two adjacent bipartite subsystems at the center, constituted of the $(N/2-1,N/2)$ and $(N/2,N/2+1)$ site pairs.  Fig.~\ref{fig:periodic} shows the variation of the quenched averaged
quantities, $\delta_{\lambda}$ (Fig.\ref{fig:periodic}(a)) and $C^{adv}_{\lambda}$, (Fig.\ref{fig:periodic}(b)) with the coupling strength $\bar{J}/h$, for both the pairs for $N=20$. We find that the results for the pairs agree with 
each other for all $\bar{J}$. The consensus of the results demonstrate that the effective environment is essentially similar for both the pairs, 
ensuring the translational symmetry
of the quenched averaged observables associated with these 
subsystems -- a fact that would naturally be followed in the case of the closed chain. In Fig. \ref{fig:scaling}, we additionally show the behavior of the quantities $\delta_{\lambda}$ and $C^{adv}_{\lambda}$ between the spins $N/2$ 
and $N/2+1$ as functions of $\bar{J}/h$ for $N=30$ and $N=50$, along with \(N=20\). 
We 
find qualitatively consistent results with the observations previously made for the random $XY$ spin models. The nearest-neighbor spin pairs at the center of the quantum $XYZ$ spin glass chain exhibit
Bell inequality violation and also turn out to 
be efficient for dense coding. We find that the changes in the quenched averaged quantities to be minimal for the different system sizes considered in Fig. \ref{fig:scaling}. 
Moreover, we observe that quenched Bell inequality violation and advantage in dense coding capacity after quenching, increase with the introduction of the \(zz\)-interaction, i.e. with the introduction of \(\Delta\).  
We specifically choose $\bar{J}=0$, where the Bell inequality 
violation and the dense coding capacity reach their respective maximum,  to illustrate the finite-size scalings 
(see Fig.~\ref{fig:scaling1}).  We find that $\delta_{\lambda, max}$ and $C^{adv}_{\lambda, max}$ decay as $N^{-1.92}$ 
and $N^{-2.24}$, respectively.

\section{Conclusion}
\label{sec:conclusion}
In this work, we have considered quenched disordered spin chains for investigating 
the monogamy of Bell inequality violation and the exclusion principle in dense coding of the three-spin nearest neighbor clusters of the corresponding zero-temperature states. 
In particular, we focus on the zero-temperature states of the random $XY$ 
spin models and the random Heisenberg spin glass. Our analysis reveals that although the monogamy of quantum correlations and the translational invariance of the Hamiltonian in clean system force the considered quantum characteristics to attain 
at most classical values -- leading to no-go theorems -- the quantum nature can be resurrected by the introduction of quenched disorder in the system.
%
The Hamiltonian itself is not translational invariant in the quenched system but the physically relevant post-quenched 
observables are so, and it is possible for the system to violate the monogamy of Bell inequality violation and quantum advantage for dense coding.
The no-go theorems are at the level of observables, and in quenched disordered systems, it is the post-quenched quantities (and not the pre-quenched ones) that are physically meaningful.  
The analysis for the disordered quantum $XY$ models is carried out by using the Jordan-Wigner, Fourier and Bogoliubov transformations, while that for the disordered quantum Heisenberg model is by employing density matrix renormalization group
techniques. Finite-size scaling analysis is performed for all the models and for both the quantum characteristics considered.

There is an ongoing effort in conquering no-go theorems in quantum mechanics either by going beyond the static framework of the quantum formalism \cite{pr-box-plus} or by relaxing quantum dynamical 
postulates like unitarity \cite{non-linear-dynamics}. The work presented in  this paper 
shows another path for 
overcoming the no-go theorems of ordered systems, while still remaining within the quantum realm, by introducing impurities or defects.

\acknowledgments
We acknowledge computations performed at the cluster computing facility at the Harish-Chandra Research Institute, India. This work has been developed by using the DMRG code released within the ``Powder with Power" project (www.dmrg.it).



\begin{thebibliography}{10}
\bibitem{vonneumann} Here, $S(\varsigma) = -\mbox{tr}(\varsigma \log_2 \varsigma)$ is the von Neumann entropy corresponding to the quantum state $\varsigma$.
\end{thebibliography}
\end{document}